# Nanostructured optical waveguide with a highly confined mode

**E. D. CHUBCHEV,[1,*] I. A. NECHEPURENKO,[1,2] A. V. DOROFEENKO,[1,2,3,4] A. P. VINOGRADOV,[1,3,4] AND ALEXANDER A. LISYANSKY[5,6]**

[1]*Dukhov Research Institute of Automatics (VNIIA), 22 Sushchevskaya, Moscow 127055, Russia*
[2]*Kotel'nikov Institute of Radioengineering and Electronics of RAS, Mokhovaya 11-7, Moscow 125009, Russia*
[3]*Institute for Theoretical and Applied Electromagnetics of Russian Academy of Sciences, 13 Izhorskaya, Moscow 125412, Russia*
[4]*Moscow Institute of Physics and Technology, Dolgoprudny, Moscow Region 141700, Russia*
[5]*Department of Physics, Queens College of the City University of New York, Queens, New York 11367, USA*
[6]*The Graduate Center of the City University of New York, New York 10016, USA*
*eugene.chubchev@yandex.ru*

**Abstract:** We propose a transmission line working at telecom wavelengths with cross section as small as $\lambda^2/39$, which is 1.6 times smaller than that of optimized silicon waveguide. The proposed line can be implemented as a subwavelength fiber with plasmonic cladding. This considerable decrease in the line cross section is achieved by utilizing a plasmonic quasi-antisymmetric mode. The required plasmonic cladding is rather thin, therefore, losses are moderate and could be compensated by using amplifying core materials. Such a transmission line can find applications in densely integrated optical systems.

## 1. Introduction

The development of modern information technologies requires increasing the clock frequency, preferably up to the optical frequencies, and reducing the mode area of transmission lines. To achieve these goals, one requires materials that allow for strong mode confinement and long propagation distances. Conventional circuits for direct or alternative currents, as well as for transmission lines operating at radio and microwave frequencies, are usually made of high conducting metals. To minimize loss, the thickness, $\delta_s$, of the skin layer should be smaller than the thickness $L_m$ of metal elements. To achieve a subwavelength cross section of a transmission line, $L_m$ must be much smaller than the free-space wavelength $\lambda_0$. Thus, for a small cross section and a long path, one needs $\lambda_0 \gg L_m \gg \delta_s$. These inequalities are easily satisfied at frequencies up to several GHz. For this frequency range, many transmission lines with small cross sections have been designed. An example of such a line is a two-wire circuit and a coaxial line.

At optical frequencies, metals are not good conductors; their skin-depth is not very small (e.g., 20–30 nm for silver). Although the optical analog of the stripline suggested in Ref. [1] has an acceptable mode area of $\sim 10^{-3}\lambda_0^2$, the size of auxiliary elements of the transmission line makes the total cross section of the line about $(\lambda_0/3)^2$. In such a line, the diameter of the high permittivity cylinder is about 400 nm, which is comparable to the typical size of a silicon waveguide [2]. Therefore, such a waveguide can hardly be coupled to electronic components, and a decrease in the waveguide cross section is still desirable.

Another approach to create optical transmission line uses the effect of total internal reflection. Transmission lines utilizing this effect are known as optical fibers [3-5]. At the telecommunication frequencies, the diameter of such waveguides is dozens of micrometers. In the surrounding medium (vacuum), the characteristic length scale of confinement of modes traveling along a waveguide with the cylindrical symmetry is determined by the imaginary part of the transverse wave number of the mode, $\kappa = \sqrt{k_z^2 - k_0^2}$, where $k_0$ is the free-space wavenumber and $k_z$ is the propagation constant. Consequently, the mode area, $S_{\text{mode}}$, is determined by $S_{\text{mode}} \sim 1/\kappa^2$. Making the cross section of the waveguide smaller than $\left(\lambda_0/\sqrt{\varepsilon}\right)^2$ results in $k_z$ approaching $k_0$, while $S_{\text{mode}} \sim 1/\kappa^2 \to \infty$.

Noticeable confinement can be achieved far from the light cone, $k_z \gg k_0$. However, even in this case, the mode area is too large because applications require optical waveguide modes with a transverse size comparable to the sizes of electronic components (dozens of nanometers) [5, 6]. The transverse size of a waveguide could be decreased by using high-permittivity materials [5]. Unfortunately, there are no transparent dielectric materials with permittivity greater than 12 [7], which is not sufficient for an acceptable reduction of the waveguide cross section [5, 8-11].

At telecom frequencies the plasmonic materials have high values of permittivity and it has been suggested several plasmonic waveguides [12-21]. Unfortunately, due to losses these waveguides, however, cannot have both a large propagation length and a small mode area.

Nevertheless, it seems that the only way out is combination of plasmonic and dielectric materials. Adding a plasmonic cladding to a dielectric waveguide leads to an appearance of two more plasmonic modes classified as $HE_1$ [22]. Plasmonic modes are generally distinguished by the distributions of the magnetic field in the cladding [23]. According to this classification, the $HE_1$ modes are referred to as quasi-symmetric and quasi-antisymmetric. In the case of the thick cladding studied in Ref. [22], the quasi-antisymmetric mode is similar to the channel mode in bulk metal, while the quasi-symmetric mode resembles the mode of a metal cylinder. In the visible range, the quasi-antisymmetric mode has a high loss that hinders its use. Therefore, in Ref. [22], the attention was focused on the quasi-symmetric $HE_1$ mode, which is localized outside the waveguide and has lower loss. The areas of $HE_{11}$ and quasi-antisymmetric $HE_1$ modes diverge as the waveguide cross section vanishes. These modes, therefore, are not acceptable for designing optical transmission lines.

In this paper, we study the quasi-antisymmetric $HE_1$ mode at the telecommunications wavelength $\lambda_0 = 1550$ nm. We show that the area of this mode decreases with a decrease in the waveguide cross section. This possibility occur because there is an interval of the waveguide radii, at which the Hashin-Shtrikman condition [24-26] for mutual subtraction of the transverse polarizations of the cladding and core is almost realized. Consequently, the fields outside the waveguide are small, and the mode area reduces to the cross section of the core. In the proposed waveguide, both the mode area and the cross section of the line are much smaller than $\lambda_0^2$. At a small thickness of the metal cladding, the mode may have a propagation length of 10 μm, which is sufficient for using it in optic-based chips.

## 2. Waveguide mode of subwavelengh area

We consider the transmission properties of a waveguide consisting of a core of a subwavelength radius $a$ and permittivity $\varepsilon_{\text{in}}$ covered with a cladding of a subwavelength thickness $d_{\text{cladd}}$ and permittivity $\varepsilon_{\text{cladd}}$ (see inset in Fig. 1). The system is surrounded by vacuum (permittivity $\varepsilon_{\text{out}} = 1$). We deal with the quasi-antisymmetric $HE_1$ mode that exists above the threshold determined by $k_0 a \sqrt{\varepsilon_{\text{in}}} \sim 1$. For the telecommunication frequencies, this condition is realized in silver, in which the real part of the permittivity of the cladding is negative ($\varepsilon'_{cladd} \ll -1$) [23]. In this mode, the field concentrates inside the core [22].

We consider the wavelength of $\lambda_0 = 1550$ nm. The values of $d_{cladd}$ are chosen to be smaller or equal to the silver skin-depth, which is about 30 nm [15]. To begin with, we assume that $a = 120$ nm and $d_{cladd} = 20$ nm. The dielectric constant of the core is $\varepsilon_{in} = 11$. The cladding material is silver with a permittivity at $\lambda_0$ is $-129+3.28i$ [27]. The wavenumber $k_z$ of the quasi-antisymmetric plasmon exceeds (Fig. 1, the intersection of the red solid dispersion curve with the horizontal line corresponding to $\lambda_0 = 1550$ nm). At this wavelength, the waveguide is practically single-mode, because the quasi-symmetric $HE_1$ plasmonic mode appears at higher frequencies at the UV range (not shown in Fig. 1). The $HE_{11}$ mode is close to the light cone (solid blue curve in Fig. 1) and almost transforms into a plane wave. It has a mode area much larger than $\lambda_0^2$. Consequently, it is required significantly more power to excite the mode.

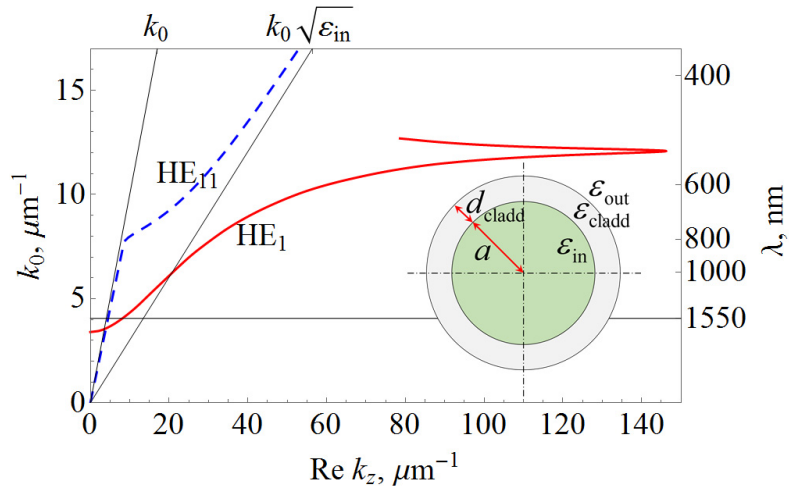

Fig. 1. Dispersion curves for the two lowest modes in the waveguide with a core radius of 120 nm core and cladding thickness of 20 nm. Straight lines show the light lines for the vacuum ($k_0 = k_z$) and the core material ($k_0 = k_z / \sqrt{\varepsilon_{in}}$). Solid red and dashed blue curves correspond to $HE_1$ (quasi-antisymmetric in metal) and $HE_{11}$ modes, respectively.

Due to the cylindrical symmetry of the problem, the transverse components of the electric $\mathbf{E}$ and magnetic $\mathbf{H}$ fields of the $HE_1$ mode are connected via the relations following from the Maxwell equations [9]

$$E_r^{(\text{out})} = -\frac{1}{\kappa^2}\left(ik_z C_E \frac{\partial}{\partial r} - \frac{k_0}{r}C_H\right)\frac{K_1(\kappa r)}{K_1(\kappa b)},$$

$$E_\varphi^{(\text{out})} = -\frac{1}{\kappa^2}\left(-ik_0 C_H \frac{\partial}{\partial r} - \frac{k_z}{r}C_E\right)\frac{K_1(\kappa r)}{K_1(\kappa b)},$$

$$H_r^{(\text{out})} = -\frac{1}{\kappa^2}\left(\varepsilon_{\text{out}} \frac{k_0}{r}C_E + ik_z C_H \frac{\partial}{\partial r}\right)\frac{K_1(\kappa r)}{K_1(\kappa b)}, \quad (1)$$

$$H_\varphi^{(\text{out})} = -\frac{1}{\kappa^2}\left(ik_0 \varepsilon_{\text{out}} C_E \frac{\partial}{\partial r} - \frac{k_z}{r}C_H\right)\frac{K_1(\kappa r)}{K_1(\kappa b)},$$

where is the external radius of the waveguide, $\varepsilon_{\text{in}}$ is permittivity of the waveguide core, $K_1$ is the modified Bessel function of the second kind. $C_E$ and $C_H$ are equal to the amplitudes of longitudinal field components $E_z^{(\text{out})}$ and $H_z^{(\text{out})}$ at the outer waveguide boundary [9, 28]. The relation between $C_E$ and $C_H$ are found from the boundary conditions at $r = a$ and $r = b$. Their absolute values are normalized by the condition $\max\left(\frac{\partial(\varepsilon\omega)}{\partial\omega}|\mathbf{E}|^2 + |\mathbf{H}|^2\right) = 1$.

Since we are interested in the modes with the subwavelength area, we have to choose both the core radius and the cladding thickness to be smaller than the free-space wavelength ($k_0 a \ll 1$ and $k_0 d_{\text{cladd}} \ll 1$). Therefore, in the neighborhood of the core-cladding waveguide, a near-field region $k_0 r \ll 1$ appears. We deal with the values of $k_z$ which are only slightly larger than $k_0$. It means that in the near-field region, an inequality $\kappa \ll k_0$ holds, resulting in $\kappa r \ll 1$. Then, in Eq. (1), the functions $K_1(\kappa r)$ may be expanded in the power series

$$K_1(\kappa r) \approx 1/(\kappa r) + (2\gamma - 1)\kappa r/4 + (\kappa r/2)\ln(\kappa r/2) + O\left((\kappa r)^2\right), \quad (2)$$

where $\gamma$ is the Euler constant.

First, for $\kappa r \ll 1$, in Eq. (2), we can retain only the first term, $K_1(\kappa r) \approx 1/(\kappa r)$. In this approximation, the derivative $\partial/\partial r$ can be replaced with $-1/r$, and the first equation in system (1) is reduced to

$$E_r^{(\text{out})} = (ik_z C_E + k_0 C_H)\frac{b}{\kappa^2 r^2}. \quad (3)$$

The other field components are reduced in the same manner.

Finally, the near-field contribution outside the core and cladding into the mode area, $S_{\text{mode}} = 2\pi \int_0^{+\infty}\left(|\mathbf{E}|^2 + |\mathbf{H}|^2\right)r\,dr$, may be estimated as

$$S_{NF} \sim \int_b^{+\infty}\left|E_r^{(\text{out})}\right|^2 r\,dr = \frac{1}{\kappa^4}|ik_z C_E + k_0 C_H|^2 b^2 \int_b^{+\infty}\frac{dr}{r^3} \sim \frac{1}{\kappa^4}|ik_z C_E + k_0 C_H|^2 \quad (4)$$

In the far-field region, $\kappa r \gg 1$, $K_1(\kappa r) \approx \sqrt{\pi/2\kappa r}\exp(-\kappa r)$. Then, from Eqs. (1) we obtain

$$E_r^{(\text{out})} \sim \frac{ik_z C_E}{\kappa}\sqrt{\frac{b}{r}}\exp(-\kappa r), \quad (5)$$

and the far-field contribution into the mode area results in

$$S_{FF} \sim \int_{1/\kappa}^{+\infty} \left|E_r^{(out)}\right|^2 rdr \approx \frac{bk_z^2}{2\kappa^3}|C_E|^2 e^{-2}.$$

High confinement requires that $S_{NF} \gg S_{FF}$. Then, $S_{mode}$ is determined by $S_{NF}$ only, and we obtain the condition for the subwavelength mode area

$$b \ll e^2 \frac{\left|(k_z/k_0)C_E - iC_H\right|^2}{\kappa|C_E|^2}. \tag{6}$$

Under this condition, the field is localized at a length scale much smaller than $1/\kappa$. Since at such distances, the field decreases considerably, the mode confinement in the suggested waveguide does not require large wavenumbers, as it is usually believed. Comparing the field distributions for different system parameters (Fig. 2), we can see that the subwavelength mode confinement can be obtained in a broad range of parameters. The key moment is that the mode area is formed in the near-field region of the waveguide. The contribution of the far-field region decays exponentially and is negligibly small (see Fig. 2).

The reason for the field suppression outside the waveguide is the Hashin-Shtrikman effect of the dielectric dipole moment compensation by a metallic cladding, which leads to the field concentration inside the core.

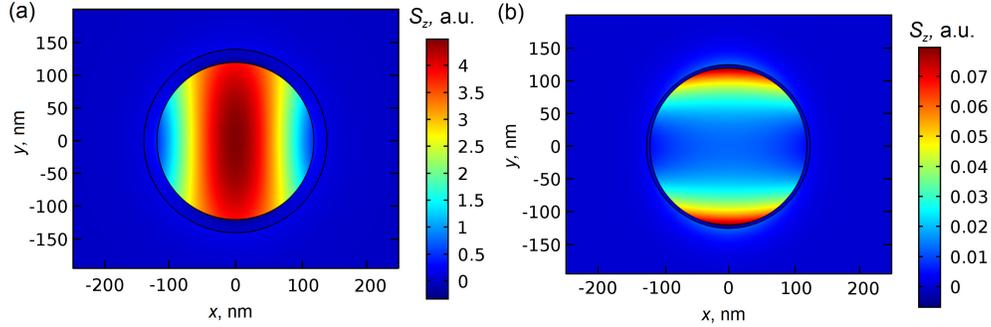

Fig. 2. The distribution of the Poynting vector of the mode at (a) $a = 120$ nm and $d_{cladd} = 20$ nm, and (b) $a = 120$ nm and $d_{cladd} = 5$ nm. The corresponding values of the propagation constant are $k_z = (2.01 + 0.029i)k_0$ and $k_z = (4.46 + 0.09i)k_0$ for (a) and (b), respectively.

In order to show that the subwavelength mode confinement can be obtained in a wide range of parameters, we show the mode area as a function of the core radius $a$ and cladding thickness $d_{cladd}$ (see Fig. 3). The mode area is calculated via the equation $S_{mode} = 2\pi \int_0^{+\infty} \left(\frac{\partial(\varepsilon\omega)}{\partial\omega}|\mathbf{E}|^2 + |\mathbf{H}|^2\right)rdr$ with the normalization condition $\max\left(\frac{\partial(\varepsilon\omega)}{\partial\omega}|\mathbf{E}|^2 + |\mathbf{H}|^2\right) = 1$.

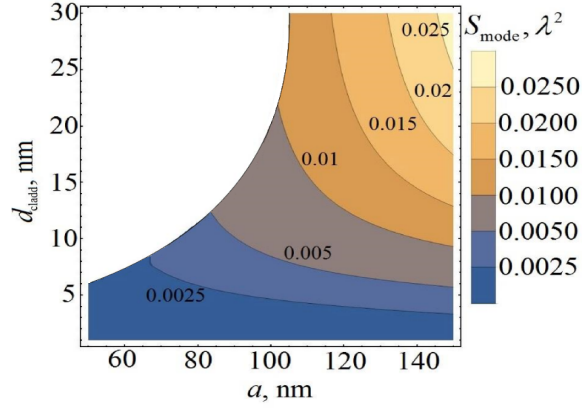

Fig. 3. The mode area as a function of the waveguide radius $a$ and coating thickness $d_{cladd}$. In the white area, the mode is leaky. The boundary of the white region corresponds to the cutoff where $k_z$ approaches $k_0$. The value of $1/\kappa$ grows near the boundary; however, the mode area stays deeply subwavelength as has been discussed above.

It should be noted that the mode area $S_{mode}$ can take values less than $\pi(a+d_{cladd})^2$. The reason for this is a concentration of the field in some part of the waveguide core. Inside the waveguide core, the fields are expressed via the Bessel function, $J_1\left(\sqrt{\varepsilon_{in}k_0^2 - k_z^2}\,r\right)$, and its derivatives. These functions oscillate at $k_z < \sqrt{\varepsilon_{in}}k_0$, and the field intensity has a maximum in the center of the core. On the other hand, at $k_z \gg \sqrt{\varepsilon_{in}}k_0$ and $k_z a \gg 1$, the Bessel function decays into the core, $J_1\left(\sqrt{\varepsilon_{in}k_0^2 - k_z^2}\,r\right) \sim \exp\left(\sqrt{k_z^2 - \varepsilon_{in}k_0^2}\,r\right)$, so that the field is concentrated in the cladding. This leads to a significant reduction of $S_{mode}$ below the waveguide cross-section $\pi a^2$ at small values of $d_{cladd}$ (see Fig. 4), at which large values of $k_z$ are reached. Indeed, at small plasmonic mode becomes electrostatic, and $k_z \sim 1/d_{cladd}$ scaling law is satisfied.

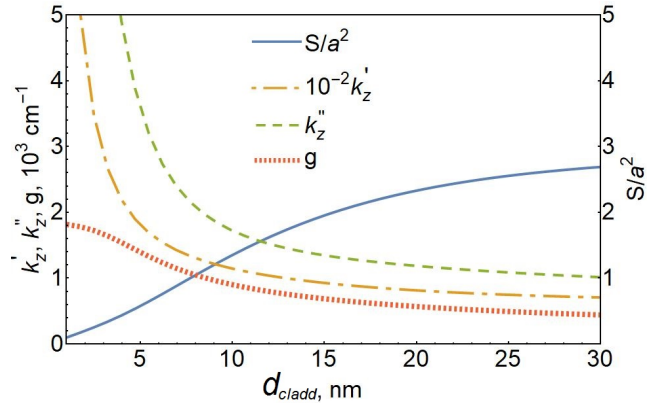

Fig. 4. The mode area $S_{mode}$ (the solid line), the real (the dash-dotted yellow line) and imaginary (the dashed green line) parts of $k_z$, and the material gain required for loss

compensation (the red dotted line) as a function of the cladding thickness $d_{cladd}$ at radius $a = 120$ nm.

## 3. Compensation for losses

Since in the optical region, all the plasmonic metals have significant Joule losses, the propagation length of the waveguide mode is not large. In Fig. 5(a), the propagation length $l_{pr} = 1/(2\,\text{Im}\,k_z)$ is shown as a function of $a$ and $d_{cladd}$. For the waveguide with the core radius of 120 nm and the cladding thickness of 20 nm, the mode propagation length is equal to $l_{pr} = 4.2$ μm, which is sufficient for future applications. Note that when the cladding thickness decreases, the volume occupied by metal also decreases. Nevertheless, due to the field concentration in the metallic cladding in the electrostatic regime, losses increase.

Energy absorption can be fully compensated if the waveguide core is made of a gain medium, such as quantum dots [29, 30], quantum wells [31], or nanorods [32]. However, even these strong gain media are not sufficient for loss compensation in the most plasmonic geometries. In our system, loss compensation is possible due to the small metal cladding thickness compared to the waveguide core [33]. Since the mode energy is mainly concentrated inside the core (see Fig. 2), gain should be localized in the same region. In the proposed waveguide, loss compensation can be achieved at realistic values of material gain. Fig. 5(b) shows the material gain sufficient for loss compensation as a function of $a$ and $d_{cladd}$. We consider the core made of a GaAsBi/GaAs nanowire. Such quantum wells have the material gain of approximately 1500 cm$^{-1}$ [34] and permittivity $\varepsilon'_{in} = 11$ [35].

As shown in Fig. 5(b), the material gain sufficient to compensate the absorption is of the order of 1000 cm$^{-1}$, which is achievable employing GaAsBi/GaAs nanowires. In particular, the material gain required for loss compensation in a waveguide with the core radius of $a = 120$ nm and the cladding thickness of $d_{cladd} = 20$ nm is 568 cm$^{-1}$ at 1550nm.

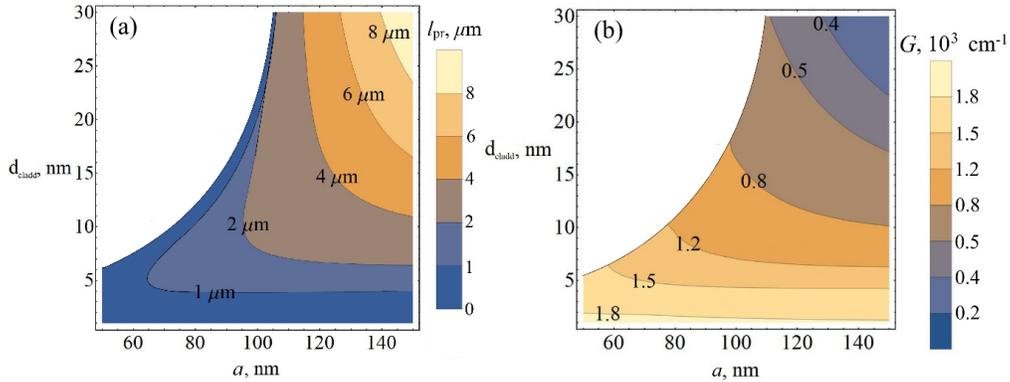

FIG 5. The propagation length (a) and the value of material gain required for loss compensation (b) vs. the radius of the waveguide and the thickness of the cladding. In the white area, the mode is leaky.

In many cases, to increase the propagation length, a partial loss compensation may be sufficient. With the help of the data for $k_z''$ in a system without gain and for the gain $G$ that provides the total loss compensation (dotted and dashed curves in Fig. 4), one can estimate the mode propagation length at any intermediate gain level $g$ ($0 < g < G$). To do this, let us note that the difference between the "material gain" $G$ and the "modal loss" $k_z''$ is due to the field distribution both inside and outside the gain medium. Let us suppose that the field fraction located in the gain medium does not depend on the gain level. Therefore, at an arbitrary gain level, the imaginary part of the wavenumber at a current level of gain, $k_z''(g)$, is a difference of its value without gain, $k_z'' = k_z''(0)$, and of the term caused by gain: $k_z''(g) = k_z'' - \chi g$. Here, the coefficient $\chi$ is calculated from the total gain compensation: $0 = k_z'' - \chi G$. Thus, one obtains $k_z''(g) = k_z'' - k_z'' g / G$, and the propagation length $L(g) = 1/2k_z''(g)$ is estimated as

$$l_{pr}(g) = \frac{l_{pr}(0)}{1 - g/G},\tag{7}$$

or $l_{pr}(g) = l_{pr}(0) G / (G - g)$. Particularly, for $l_{pr}(0) = 4.2$ μm and $G = 568\,\text{cm}^{-1}$, the value of $g = 330\,\text{cm}^{-1}$ gain increases the propagation length up to $l_{pr}(g) = 10$ μm.

## 4. Conclusion

We study a structure of an optical dielectric waveguide with a thin metal cladding. We demonstrate that utilizing the quasi-antisymmetric mode $HE_1$ allows one to construct a transmission line with a subwavelength cross section and a sufficiently long propagation length of $4.2$ μm, which is enough for the communication between usual (electronic) kernels.

In a subwavelength neighborhood of the waveguide, the electromagnetic field acquires a near-field character. This special property is accompanied by a power-law decay in 2D and 3D systems (cladding fiber and chain of spherical particles), and cannot be realized in layered systems in which there is no near-field decay. The contribution of the near-field region is crucial in the formation of the area of the mode propagating along the waveguide. Upon exiting this region, the field acquires a far-field character. In the far-field region, the field decays exponentially, $\sim \exp(-\kappa r)$, but with a very small decay rate, $\kappa = \sqrt{k_z^2 - k_0^2}$ with $k_z \sim k_0$. Therefore, strong decay in the near field makes the contribution of the far-field negligible, so that the mode area is $S_{\text{mode}} \ll 1/\kappa^2$.

In Ref. 21, a transmission line operating the quasi-symmetric mode was considered in the visible range. Although this transmission line realization has a cross section of $(\lambda_0/3)^2$, the mode area is determined by the same relation as discussed above, $S_{\text{mode}} \sim 1/\kappa^2$. Due to high loss in metal, possible values of $k_z$ are about $k_0$ [15, 36, 37] that do not allow for subwavelength confinement of the mode [38].

The range of geometrical and material parameters needed for the realization of the proposed waveguide is shown in Fig. 3. A mode area as small as $0.01\lambda^2$ at $\lambda = 1550$ nm is achievable even if the mode wavenumber is only slightly larger than the free-space wavenumber. Due to the thin cladding, Joule losses in metal can be compensated if the core of the waveguide is made of a gain medium such as quantum dots, quantum wells, or nanorods. The proposed waveguide can find applications in dense integrated optical circuits and crosstalk suppression.

**Funding**

Office of Naval Research (ONR) (N00014-20-1-2198).

**Acknowledgements**

We are thankful to Oleg Kotov and Sergey Bankov for fruitful discussions.